\documentclass{article}

\usepackage{graphicx}  
\usepackage{amsmath}   
\usepackage{amssymb}   
\usepackage{bm}

\def\eq#1{{Eq.~(\ref{#1})}}

\def\frab#1#2{\left(\frac{#1}{#2}\right)} 

\def\cN{{N(x_2,x_1;\sigma)}}

\def\cG{{G(x_2,x_1;m)}}

\def\cK{{K_0(x_2,x_1;s)}}

\def\ket#1{|#1\rangle}                    
\def\bk#1#2#3{{\langle #1|#2|#3\rangle}}  

  \title{A Measure for Quantum Paths, Gravity and Spacetime Microstructure}
  \author{T. Padmanabhan\\
  IUCAA, Pune University Campus,\\
  Ganeshkhind, Pune- 411 007.\\
  {\small {email: paddy@iucaa.in}}
  }

  \date{}

\begin{document}
  
  \maketitle
  
  \begin{abstract}
The number of \textit{classical} paths of a given length,  connecting any two events  in a (pseudo) Riemannian spacetime is, of course,  infinite.  It is, however, possible to define a useful, finite,  measure $\cN$
for the effective number of \textit{quantum} paths [of length $\sigma$ connecting two events $(x_1,x_2)$]  in an arbitrary spacetime.
 When $x_2=x_1$, this reduces to $C(x,\sigma)$ giving the measure for closed quantum loops of length $\sigma$ containing an event $x$. Both $\cN$ and $C(x,\sigma)$ are  well-defined  and depend only on the geometry of the spacetime. Various other physical quantities like, for e.g., the effective Lagrangian, can be expressed in terms of $\cN$. The corresponding measure for the total path length contributed by the closed loops, in a spacetime region $\mathcal{V}$, is given by the integral of $L(\sigma;x) \equiv \sigma C(\sigma;x)$ over $\mathcal{V}$. Remarkably enough $L(0;x) \propto R(x)$, the Ricci scalar; \textit{i.e, the measure for the total length contributed by infinitesimal closed loops in a region of spacetime gives us the Einstein-Hilbert action.}
 Its variation, when we vary the metric, can provide a new route towards induced/emergent gravity descriptions. In the presence of a background electromagnetic field, the corresponding expressions for $\cN$ and $C(x,\sigma)$ can be related to the holonomies of the field. 
The measure  $\cN$ can also be used to evaluate a wide class of path integrals for which the action and the measure are \textit{arbitrary} functions of the path length. As an example, I compute  a modified path integral which incorporates the zero-point-length in the spacetime. I also describe several other properties of  $\cN$ and outline a few simple applications.

 \end{abstract}
 
 \section{Introduction}
 
 I define and explore the properties of a function $\cN$ which I call the path measure. It can be thought of as the amplitude for a quantum path of length $\sigma$ to exist between two events $x_1$ and $x_2$ in arbitrary, curved spacetime. Mathematically it contains the same amount of information as the Feynman propagator for a massive scalar field in the spacetime or --- equivalently --- the Schwinger (heat) kernel. Conceptually, however, it is very different and is more readily amenable for studying the quantum microstructure of the spacetime. It also has better convergence properties in the coincidence limit. 
 
 The main purpose of this paper is to introduce this quantity, obtain some of its key properties and describe a few applications, postponing more detailed discussion to   future publications. In particular, I will show here (i) how the Einstein-Hilbert action has a natural representation in terms of the path density and (ii) how a large class of relativistic path integrals can be evaluated using this concept.

 \section{Counting the uncountable}
 
 Consider a scalar field $\phi(x)$ which satisfies the Klein-Gordon equation
 $(\Box + m^2)\phi(x) =0$ in an arbitrary spacetime with metric $g_{ab}$; the  $\Box$ is the Laplacian corresponding to this metric.\footnote{I use the signature $(+ - - -)$ and work with dimensionless quantities by rescaling \textit{all} the variables by suitable powers of a length scale $\ell_P$ defined as $\ell_P\equiv\lambda(G\hbar/c^3)^{1/2}$, where $\lambda$ is a numerical factor introduced for future convenience. The units are chosen making $\ell_P=1$ so that it need not be displayed.
  Latin indices run over the spacetime coordinates while the Greek indices run over spatial coordinates. I will write just $x$ for $x^i$ etc. when no confusion is likely to arise.  Analytic continuation to Euclidean sector will be used, whenever it is appropriate, to make the expressions unambiguous.} The Feynman propagator for this field can be expressed in terms of the Schwinger proper time kernel in the form:
 \begin{equation}
 G(x_2,x_1; m) = \int_0^\infty ds\ e^{-im^2 s } \ K_{m=0} (x_2,x_1; s)
  \label{one}
 \end{equation}
 where 
 \begin{equation}
  K_0 (x_2,x_1; s)  \equiv  K_{m=0} (x_2,x_1; s)  = \langle x_2|e^{-is\Box}|x_1\rangle
  \label{two}
 \end{equation}
 is the kernel for the massless field. This quantity is  independent of the properties of the scalar field and is a purely  geometrical object determined by the metric of the background spacetime.\footnote{To be precise, it also depends on the boundary conditions which will choose the vacuum states used to define the propagator. These boundary conditions will not play any  role in most of our discussions and we will discuss this dependence in Sec. \ref{rindler}. Also note that the convergence of the integral in \eq{one} requires the prescription $m^2 \to m^2 -i\epsilon$; therefore we can think of $K_0$ as carrying a regulator $e^{-\epsilon s}$ in it. I will not bother to indicate explicitly such convergence factors for integrals except when it is vital to the discussion.}  
 
 The Feynman propagator can also be given a formal path integral definition in the form 
 \begin{equation}
  G(x_2,x_1; m) = \sum_{\rm paths} \exp -im\ell (x_2,x_1)
  \label{three}
 \end{equation}
 where $\ell(x_2,x_1)$ is the length of the path connecting the two events $x_2, x_1$ and the sum is over all paths connecting these two events. Recall that I have made all quantities dimensionless and hence $G$ in \eq{three} actually stands for $\ell_P^{2}G_{\rm conv}$ where $G_{\rm conv}$ is the conventional propagator with the dimension of (length)$^{-2}$. This will make the measure in \eq{three} dimensionless. (From \eq{one} we know that $G(x_2,x_1; -m) = G(x_2,x_1;m)$. So we could have also written \eq{three} as a sum over $(1/2) [e^{-im\ell} + e^{im\ell}]$. This fact will prove to be useful at a later stage; for now, I will continue to work with the sum in \eq{three}.)
 The definition is formal because I have not specified the measure to be used in the path integral.\footnote{This is, in general, a non-trivial issue. In flat spacetime, one can  define the path integral in a Euclidean cubic lattice, take the continuum limit with suitable measure and analytically continue the result to Minkowski spacetime (see p. 41 of \cite{tpqft});
 but there is no simple generalization to curved spacetime.}
 
 I will now describe another way of giving a physical meaning to the above path integral. To do this, note that with a judicious introduction of Dirac delta function, \eq{three} can be re-written in the form:
 \begin{equation}
 G(x_2,x_1; m) = \int_{-\infty}^\infty d\sigma\ e^{-im\sigma} \sum_{\rm paths}\delta_D \left(\sigma - \ell (x_2,x_1)\right)
 \equiv \int_{-\infty}^\infty d\sigma\ e^{-im\sigma} N(x_2,x_1; \sigma)
  \label{four}
 \end{equation}
 where we have defined the function $N(x_2,x_1; \sigma)$ to be:
 \begin{equation}
 N(x_2,x_1; \sigma) \equiv \sum_{\rm paths}\delta_D \left(\sigma - \ell (x_2,x_1)\right)
  \label{five}
 \end{equation}
 Obviously, $\cN$ can be thought of as a quantum measure ``counting'' the ``number'' of paths of length $\sigma$ between the two events $x_2,x_1$. We will see soon that $\cN$ is finite and well defined even though the right hand side of \eq{five} is made of sums of Dirac delta functions. The path integral measure is \textit{defined} such that the sum over the \textit{quantum} paths leads to the correct propagator, given by \eq{one}. 
  In fact, \eq{four} tells us that the propagator for the zero mass particle is given by:
 \begin{equation}
 G(x_2,x_1; 0) =  \sum_{\rm paths}\ 1
 = \int_{-\infty}^\infty d\sigma\  N(x_2,x_1; \sigma)
  \label{tpfour}
 \end{equation}
 which is finite with the path integral measure defined as described. 
 
 The measure  $\cN$ allows us to convert path integration to ordinary integration. It is clear that:
\begin{equation}
 \sum_{\rm paths} \exp -im\ell (x_2,x_1) =  \int_{-\infty}^\infty d\sigma\ e^{-im\sigma} N(x_2,x_1; \sigma)                                    
 \end{equation}
 So specifying $\cN$ is equivalent to specifying the path integral measure and it contains all the information we need to determine the propagator. More explicitly,
 inverting the Fourier transform in \eq{four} we can  find an expression for $\cN$ in terms of the Feynman propagator $\cG$. We get\footnote{The introduction of the Dirac delta function in the path integral uses integration of $\sigma$ over real line and assumes $\ell$ to be real. For spacelike section of paths $\ell$ can become (formally) complex. It turns out that this does not create any real problem because the path integral has to be \textit{computed} either by rewriting it in terms of the squared path length or in a Euclidean lattice \cite{tpqft}. The Fourier relations in \eq{four} and \eq{six} can be taken as the basic definition of $N(x_2,x_1; \sigma)$ though the motivation is clearer in the path integral approach.}
 \begin{equation}
 N(x_2,x_1; \sigma) = \int_{-\infty}^\infty \frac{dm}{2\pi} \ \cG \ e^{im\sigma}
  \label{six}
 \end{equation}
 Here and in what follows we are treating $m$ purely as a parameter. It is obvious from \eq{one} that the propagator $\cG$ depends on $m$ only through $m^2$. This allows extending the definition of $\cG$ to negative values of $m$ by using the fact that $\cG$ is an even function of $m$. While performing the integral in \eq{six} we can also work with Euclidean spacetime (and the Euclidean propagator $G_E$) thereby defining a Euclidean version of $N_E$; analytic continuation will then give an appropriate expression in the Minkowskian spacetime. This procedure will also take care of the issue of boundary condition, mentioned earlier, to a great extent.\footnote{It is convenient to use the Fourier transform with the factor $e^{im\sigma}$ both in the Euclidean and Lorentzian sectors.}
 
 The expression in the right hand side of \eq{four} has  a heuristic physical interpretation. When a particle propagates between two events $x_2$ and $x_1$ along a timelike curve of length $\sigma$, the probability amplitude will acquire a phase $\exp (-im\sigma)$ because we can think of $m$ as the energy in the rest frame of the particle. If quantum amplitude for a path of length $\sigma$ to exist between the two events is $\cN$, then the total amplitude for the particle to propagate from $x_1$ to $x_2$ is clearly given by the expression in the final integral in \eq{four}. This generalizes the naive notion of ``counting'' the number of paths of a given length between two events. The  paths  which actually contribute to the path integral  are, of course,  not  smooth, classical, geometrical paths.  It is therefore more useful to think of  $\cN$  as a \textit{quantum} measure in the space of paths, viz. quantum amplitude for a path of length $\sigma$ to exist between the two events, thereby generalizing the notion of classical ``counting'' to quantum path integral. 
 
 We will see that this expression $\cN$ in \eq{six} --- which gives the  quantum measure for the  paths which \textit{actually} contribute to the path integral --- is finite, in sharp contrast to the number of classical smooth paths, which, of course, is divergent.
 In the Euclidean sector, $\cG$ is real and an even function of  $m$, making $\cN$ real. In the Lorentzian sector, because 
 $\cG$ is complex,  $\cN$ will be complex, again emphasizing the quantum nature of the measure; most of the time we will do the calculations in the Euclidean sector.
 
 Given $\cN$ we can define number of closed loops of length $\sigma$, starting and ending at an event $x$, by
  setting $x_2=x_1=x$ in $\cN$.  This leads to:
 \begin{equation}
  C(\sigma;x)\equiv N(\sigma;x,x)=\lim_{x_2\to x_1}\int_{-\infty}^{+\infty}\frac{dm}{2\pi} \ \ \cG \ e^{im\sigma}
  \label{newseven}
 \end{equation}
 Note that we first evaluate the integral and then take the coincidence limit. This will lead to a finite quantity even though the coincidence limit of the propagator itself is divergent.\footnote{This is just a trivial consequence of the fact that while $C(\sigma;x)$ is finite and well-defined, its Fourier transform with respect to $\sigma$ --- which gives $G(x,x)$ --- does not exist.}
 
  The relation between $\cG$ and $\cN$ is very similar to the relation between $\cG$ and $K(x_2,x_1,s)$, expressed by \eq{one}, with two key differences. 
 
 (i) The limits of integration for $s$ in \eq{one} are zero and infinity while  $\sigma$ is integrated from $-\infty$ to $+\infty$ in \eq{four}. This can, however, be easily taken care of by extending the integration range in \eq{one} to the entire real line and inserting a $\theta(s)$ factor in the integral; i.e., by modifying $\cK$ to $\theta(s) \cK$. This modification can even be interpreted as due to the particle always propagating forward in Schwinger proper-time. 
 
 (ii) The more crucial issue is  that the phase factor in \eq{one} has $m^2$ in it (which is hard to interpret) while the phase factor in \eq{four}  has only $m$ (which is physically understandable). If we rescale $s\to s/m$, the phase factor can be readjusted in \eq{one} but this will change $\cK$ to $m^{-1} K_0(x_2,x_1; s/m)$. The new kernel now depends on $m$ and this dependence could be non-trivial in an arbitrary curved spacetime. This is not desirable since we would like the heat kernel to be a purely geometrical object depending only on the properties of the Laplacian, $\Box$, in a given spacetime. 
 
 Except for these two crucial differences,  $\cN$ and $\cK$ are very similar in structure, and we will see that  $\cN$ has better convergence properties and  a direct physical interpretation (as the quantum measure for paths of a given length).

 \section{Path measure in flat spacetime}
 
 In the rest of the paper, we will explore various properties and applications of the effective number of paths $\cN$.  In general, we will work in an arbitrary (non-flat) spacetime. However, before we do that, in this section, let me exhibit the form of $\cN$ in the \textit{flat} spacetime both in the Lorentzian and Euclidean sectors. An elementary calculation (see Appendix 1; also see \cite{kkap}, Eq (411)) shows that in this case 
 \begin{equation}
 \cN  = 
 \begin{cases}
   \frac{1}{4\pi} \frac{\Gamma(k)}{\pi^k} \frac{1}{(\sigma^2 + x^2 )^k} &\text{(Euclidean)}\\
    \frac{1}{4\pi} \frac{\Gamma(k)}{\pi^k} \frac{1}{(\sigma^2 - x^2 +i \epsilon)^k} &\text{(Lorentzian)}
 \end{cases}
  \label{apseven}
 \end{equation}
 where $k= (1/2)(D-1)$ and $x^2=(x_1-x_2)^2$. As advertised, it is  well defined and in fact relatively simple in form. (The Euclidean measure is everywhere finite while the Lorentzian one has singularities on the light cone, reminiscent of the propagator.) One can also verify that: (i) The expression in the Lorentzian sector arises from the analytic continuation of the one in the Euclidean sector. (ii) Using these expressions in \eq{tpfour} leads to the correct propagator for the massless case. 
 
 It is possible to understand the result in \eq{apseven}  from a simple argument. We know that the propagator, in $D$ spacetime dimensions, satisfies the equation $(\Box_D +m^2)G=\delta(x)$. It is then easy to show that $\cN$ satisfies the differential equation 
 \begin{equation}
  \left(\Box_D -\frac{\partial^2}{\partial\sigma^2}\right)N=\Box_{D+1}N=\delta(x)\delta(\sigma) 
  \label{coulomb}
 \end{equation} 
 where $\Box_{D+1}$ is the Laplacian in a fictitious space with one extra spatial dimension having the line element $dL^2=ds^2-d\sigma^2 =g_{ab}dx^adx^b-d\sigma^2$. It follows from \eq{coulomb} that, $N$ for $D$ dimension is essentially the massless propagator in $(D+1)$ dimensional space with suitable signature. In the Euclidean sector, $N$ is just the Coulomb field of a unit charge at origin. In a \textit{flat} Euclidean space with $(D+1)$ dimensions such a field will fall as $[(D+1)-2]$ th power of the radial distance. That gives the scaling:
 \begin{equation}
  N\sim \frac{1}{(\sqrt{x^2+\sigma^2})^{D-1}}
 \end{equation} 
 which is exactly what we see from \eq{apseven}.

 The corresponding measure for closed loops of length $\sigma$ passing through any event is given by the expression:
 \begin{equation}
 C(\sigma)  = 
 \begin{cases}
   \frac{1}{4\pi} \frac{\Gamma(k)}{\pi^k} \frac{1}{(\sigma)^{D-1}} &\text{(Euclidean)}\\
    \frac{1}{4\pi} \frac{\Gamma(k)}{\pi^k} \frac{1}{(\sigma^2 +i \epsilon)^k} &\text{(Lorentzian)}
 \end{cases}
  \label{apeight}
 \end{equation}
 This is also finite and well-defined even though the propagator diverges at the coincidence limit. In the flat space(time), the coincidence limit in the \textit{Euclidean} sector identifies the two end points
 because $(x_2-x_1)^2=0$ implies $x_2=x_1$. But in the \textit{Lorentzian} sector one can also satisfy the condition 
 $(x_2-x_1)^2=0$ when the two events are connected by a null ray. This is analogous to the fact that while the Euclidean propagator diverges only in the coincidence limit, the Lorentzian propagator diverges on the light cone.
 
 Incidentally, our result implies that the measure for the path integral, for the relativistic particle, \textit{must} be defined through the sum given by 
 \begin{equation}
  \sum_{\rm paths} \delta_D(\sigma - \ell) =  \frac{1}{4\pi} \frac{\Gamma(k)}{\pi^k} \frac{1}{(\sigma^2 - x^2 +i \epsilon)^k}
  \label{apten}
 \end{equation}
 if the sum over paths in \eq{three} has to reproduce the propagator in \eq{one}.
  Somewhat surprisingly, such a simple characterization of the path integral measure does not seem to have been noticed in the literature before.

  \section{General properties of $\cN$}
  
  In this section I will describe several properties of the path measure $\cN$, in an arbitrary curved spacetime. Some of the technical details can be found in Appendix 1.
  
  \subsection{Relation to the Schwinger kernel}
  
 Let us begin by noticing that if we express $\cG$ in the form of \eq{one} we can perform the integral over $m$ appearing in \eq{six} and obtain the result
  \begin{equation}
  \cN = \int_0^\infty ds \left(\frac{1}{4\pi i s}\right)^{1/2} \ e^{i\sigma^2/4s} \ K_0 (x_2,x_1;s)
  \label{seven}
 \end{equation}
 Since $\cK$ is essentially determined by geometry of  the spacetime (for specified boundary conditions), it follows that $\cN$ is also a purely geometrical quantity.
 One way to ensure correct boundary condition is to use the Euclidean version of \eq{seven}, given by:
 \begin{equation}
  N^E(x_2,x_1;\sigma) = \int_0^\infty ds \frab{1}{4\pi s}^{1/2}\, e^{-\sigma^2/4s} \, K_0^E(x_2,x_1; s)
  \label{tp1}
 \end{equation} 
 where $K_0^E$ is the Euclidean kernel. It is clear that the convergence near $s=0$ is improved due to the factor $\exp-(\sigma^2/4s)$. So while the coincidence limit of the propagator can be divergent, the coincidence limit of $\cN$ is finite in all reasonable cases. This is another reason to use $\cN$ rather than $\cG$ though both contain the same amount of information.
 
 These equations, \eq{seven} and \eq{tp1}, reiterate the connection between $\cN$ in a D-dimensional spacetime (with the line element $ds^2=g_{ab}dx^adx^b$) and the massless propagator in $(D+1)$ dimensional spacetime (with the line element $ds^2=g_{ab}dx^adx^b-d\sigma^2$) with an extra ``flat'' spatial direction with the coordinate $\sigma$. The integrand in \eq{seven} (or \eq{tp1}) is just the zero-mass kernel in such a $(D+1)$ dimensional spacetime. Therefore $\cN$ in the D-dimensional spacetime is indeed the massless propagator $G_{m=0}(x_1,x_2;0,\sigma)$ in the $(D+1)$ dimensional spacetime (with the line element $ds^2=g_{ab}dx^adx^b-d\sigma^2$). 
 This is yet another way of understanding the result in \eq{apseven}. In fact, this feature allows the calculation of $\cN$ whenever, the $K_0^E$ has dependence through a factor  $\exp(-r^2/4s)$ on one or more coordinates with arbitrary dependence on all other coordinates. I will discuss one such example later on in Sec. \ref{rindler}.
 
 \subsection{A recurrence relation across dimensions}
 
 A closely related result corresponds to a recurrence relation across the dimensions for $\cN$.
 Consider a class of `separable' spacetimes for which the line element is `flat' in $k$ of the $D$ spatial coordinates, with the form:
 \begin{equation}
ds^2 = h_{ab} \ dx^a\, dx^b  - \delta_{ab}\ dy^a \, dy^b
  \label{fri4}
 \end{equation}
 where, temporarily, the indices on the  coordinates $x^a$ run  over $(D-k)$ coordinates while those on the (spatial) coordinates $y^a$ run over the $k$ flat coordinates. 
 I will derive a useful recurrence relation  relation between $\cN$ in the $(D-k)$ dimensional spacetime and the one in the $D$ dimensional spacetime. 
 
 I will first re-derive a recurrence relation for the \textit{propagator} (which is known in the literature \cite{recur}) and translate it into a recurrence relation for the path measure.
 To obtain recurrence relation for the propagator, we note that, in the equation satisfied by the full propagator, we can separate out $\Box_D$ into the form:
 \begin{equation}
 (\Box_D + m^2) G = (\Box_x - \delta^{ab} \frac{\partial}{\partial y^a} \, \frac{\partial}{\partial y^b} +m^2) G
  \label{fri5}
 \end{equation}
 It follows that the Fourier transform $\tilde G$ of this propagator in the $y-$coordinates will satisfy the equation $(\Box_x +m^2+p^2)\tilde G=\delta$. But this is just the equation for the propagator
  $G_{D-k}(\mu)$ for the $D-k$ dimensional spacetime (excluding  the $k$ flat directions) corresponding to the  mass $\mu^2=(m^2+p^2)$.
 It immediately follows that:
 \begin{equation}
 G_D(x_2,x_1,y;m) = \int \frac{d^k q}{(2\pi)^k} \ e^{iq\cdot y} G_{D-k} (x_2, x_1; \sqrt{q^2 + m^2})
  \label{fri6}
 \end{equation}
 Expressing the propagators in terms of the path measure, we can obtain the following recurrence relation for the latter:
 \begin{equation}
  N_D (x_2, x_1; y; \sigma) = \int_{-\infty}^\infty d\sigma_1\ N_{D-k}(x_2, x_1; \sigma_1)\  F(y; \sigma; \sigma_1)
  \label{fri7}
 \end{equation}
 where $F$ is a $k-$ dimensional Fourier transform of an integrand involving a modified Bessel function:
 \begin{equation}
  F(y; \sigma; \sigma_1) = \int \frac{d^k q}{(2\pi)^k} \ e^{iq\cdot y}  \frac{iq\sigma_1}{\sqrt{q^2-\sigma_1^2}} \ K_1\left[ q \sqrt{\sigma^2 - \sigma_1^2}\right]
  \label{fri8}
 \end{equation}
 with $\sigma^2$ interpreted as $\sigma^2+i\epsilon$ for convergence. To arrive at this result we have used the following cosine transform:
 \begin{equation}
 \int_0^\infty dx\ (\cos xy) \ e^{-\beta\sqrt{x^2 + \alpha^2}} = \frac{\alpha \beta}{\sqrt{\alpha^2 + \beta^2}} \, K_1\left( \alpha (y^2+\beta^2)^{1/2}\right)
  \label{fri9}
 \end{equation}
 valid for Re $\alpha>0$, Re $\beta>0$ which we need to impose by correct $i\epsilon$ prescription. This result is useful when the propagator for a lower dimensional spacetime is known and the additional directions are flat, like for e.g., in the case of $D$ dimensional Rindler spacetime. 
 
 \subsection{Path measure and the choice of vacuum state}\label{rindler}
 
 One can also describe the inequivalent vacua in the spacetime in terms of how they modify the path measure. I will illustrate the idea using the concept of Rindler vacuum in flat spacetime. 
 
 We know that the propagator in \eq{one} also has a representation as the vacuum correlator $\bk{M}{T[\phi(x_1)\phi(x_2)]}{M}$ where $\ket{M}$ is the usual inertial vacuum in the Minkowski spacetime. If we introduce standard Rindler coordinates in the right $(R)$ and left $(L)$ wedges of the $t-x$ plane through $x=\pm \rho\cosh \tau, t=\pm \rho \sinh \tau$, then one can quantize the scalar field in terms of positive/negative frequency modes with respective to $\tau$, in R and L. The corresponding creation and annihilation operators allow us to define the Rindler vacuum $\ket{R}$. This, in turn, allows us to define the Rindler propagator $G_R \equiv \bk{R}{T_\tau[\phi(x_1)\phi(x_2)]}{R}$ with time ordering defined in terms of $\tau$. 
 
 Since this propagator is different from the inertial propagator $\cG$ defined in \eq{one}, it follows that both the zero mass kernel $\cK$ and the path integral measure in \eq{three} has to be different (from the inertial vacuum expressions) to give the Rindler propagator $G_R$. In our approach, we can easily incorporate this effect by a modified path measure $N_R$ corresponding to the Rindler vacuum.
 This quantity can again be computed either by using \eq{six} or \eq{seven} from the known expressions for $G_R$ and $K_R$. The resulting  expression is not very illuminating since $G_R$ and $K_R$ are complicated functions \cite{rindlervac}. 
 
 There are, however, couple of general results one can immediately obtain for $N_R$ from the fact that Rindler vacuum contains a thermal population of Minkowski particles. First, it is well known that the Minkowski and Rindler propagators are related by the thermalization condition:
  \begin{equation}
   G_M (\tau) = \sum_{n=-\infty}^\infty G_R (i\tau + 2\pi n)
   \label{thermaltp}
  \end{equation} 
  This essentially encodes the KMS condition through the periodicity in imaginary time. Since $N_M$ and $N_R$  are defined through integrals like the one in \eq{six}, it immediately follows that the path measure also satisfies the thermalization condition given by
  \begin{equation}
   N_M (\tau) = \sum_{n=-\infty}^\infty N_R (i\tau + 2\pi n)
   \label{thermaltp1}
  \end{equation}
 In Ref. \cite{rindlervac} it has been shown how this result can be inverted through an integral kernel which allows the explicit computation of $N_R$ from the known form of $N_M$. 
 
 Second, the result obtained in \eq{fri7} --- which is valid in an arbitrary curved spacetime as long as there are some flat directions --- can be simplified further in the case of Rindler vacuum. To do this, we note that  $K_R(\Delta \mathbf{x}_\perp,\rho_1, \rho_2, \tau_1,\tau_2;s)$ and $G_R(\Delta \mathbf{x}_\perp,\rho_1, \rho_2, \tau_1,\tau_2)$ depend only on the difference $\Delta \mathbf{x}_\perp\equiv (\bm{x}^\perp_1 -\bm{x}^\perp_2)$  in the $(D-2)$ transverse coordinates. One can eliminate this dependence on the transverse coordinates by Fourier transforming $G_R$ (or $K_R$) on the transverse coordinates thereby introducing a conjugate variable $\bm{k}$. The Fourier transform $\bar G_R (\bm{k}, \rho_1, \rho_2, \tau_1,\tau_2) $ has a universal behaviour independent of the dimension $D$; the $k$-dependence  arises only through the replacement of $m^2$ by $\mu^2 \equiv m^2 + k^2$. 
 On the other hand, \eq{seven} tells us that the expression for $D$-dimensional path measure is the same as the expression for $D+1$ dimensional zero mass propagator with 
 the replacement of $(\bm{x}^\perp_1 -\bm{x}^\perp_2)^2$ by $(\bm{x}^\perp_1 -\bm{x}^\perp_2)^2+ \sigma^2$. Therefore one can compute the path measure for Rindler vacuum from the known expressions for the massless propagator in one higher dimension.
 
 Algebraically, this ides works out as follows. The Rindler propagator in $D$ dimensions, when expressed as an integral over the Rindler Kernel, has the form:
 \begin{equation}
  G_R=\int_0^\infty ds \left(\frac{1}{4\pi i s}\right)^{D/2}F(\rho_1,\rho_2,\tau;s)
  \exp\left[-im^2s -\frac{i}{4s}(\Delta \mathbf{x}_\perp)^2\right]
  \label{may201}
 \end{equation}
 where $\tau=\tau_2-\tau_1$ is the difference in the Rindler time between the two events. This form arises from the fact that the Gaussian dependence of the kernel in the transverse coordinates separates out. (One can, of course, compute $F(\rho_1,\rho_2,\tau;s)$ --- see e.g. Ref. \cite{rindlervac} --- but we will not need that result.) Using \eq{six} we will now get the path density in a form analogous to \eq{seven}:
 \begin{equation}
  N_R=\int_0^\infty ds \left(\frac{1}{4\pi i s}\right)^{(D+1)/2}F(\rho_1,\rho_2,\tau;s)
  \exp\left[ -\frac{i}{4s}[(\Delta \mathbf{x}_\perp)^2+\sigma^2]\right]
 \end{equation} 
This is exactly the massless propagator in $D+1$ dimension, obtained from \eq{may201} by setting $m=0$, replacing $D$ by $(D=1)$
and adding an extra `coordinate' $\sigma$.
The result simplifies somewhat if we eliminate the transverse coordinates by Fourier transforming both sides of \eq{may201} with respect to $\Delta \mathbf{x}_\perp$, thereby introducing a conjugate variable $\bm{k}$. Such a Fourier transform changes $m^2$ to $\mu^2\equiv m^2+k^2$. Then, $G_R$ can be expressed in the form:
\begin{equation}
 G_R=\int \frac{d^{D-2}k}{(2\pi)^{(D-2)}}\bar G(k^2+m^2;\rho_1,\rho_2,\tau)\exp (i\mathbf{k}\cdot \Delta \mathbf{x}_\perp) 
\end{equation} 
where the over-bar denotes the function in Fourier space.
Clearly, the corresponding Fourier transform of $N_R$ with respect to $\Delta \mathbf{x}_\perp$ will be:
\begin{equation}
 \bar N_R=\int_{-\infty}^{+\infty} \frac{dq}{(2\pi)}\bar G(k^2+q^2;\rho_1,\rho_2,\tau)\exp (iq\sigma) 
\end{equation}
which, of course, is the zero-mass propagator in $D+1$ dimension, Fourier transformed with respect to the transverse coordinates.
This allows us to compute $N_R$ in $D-$ dimensional Rindler vacuum from the known forms (see e.g. \cite{recur}) of massless Rindler propagator in $(D+1)$ dimension.

\subsection{Effect of spacetime curvature on the path measure}
  
  The form of $\cN$, of course, depends on the background curvature.
We can quantify the dependence of $\cN$ on the spacetime geometry, fairly generally, along the following lines. As is well-known, the Schwinger kernel in an arbitrary spacetime has an expansion in terms of Schwinger-DeWitt coefficients \cite{sch-dew} which allows us to write
 \begin{equation}
K_0(x_2,x_1; \sigma) = \frac{\Delta^{1/2}(x_2,x_1)}{(4\pi is)^{D/2}}\, e^{i\rho^2(x_2,x_1)/4s} F(x_2,x_1;s)
  \label{mon1}
 \end{equation} 
 where
 \begin{equation}
 \rho(x_2,x_1) 
= \int_0^s ds' \left\{g_{\mu\nu}\frac{dx^{\mu}}{ds'}
\frac{dx^{\nu}}{ds'}\right\}^{1/2}
  \label{prd138}
 \end{equation}
 is the geodesic distance, $\Delta(x_2,x_1)$ is the Van-Vleck determinant:
 \begin{equation}
 \Delta(x_2,x_1) 
= \Biggl(-[-g(x)]^{-1/2}\; 
{\rm det}[\partial_{\mu}^{(1)}\partial_{\nu}^{(2)}\rho(x_2,x_1)]
\left[-g(x')\right]^{-1/2}\Biggl).
  \label{prd139}
 \end{equation}
and $F$ has the asymptotic expansion in the form: 
 \begin{equation}
 F(x_2,x_1;s) 
= \sum_{n=0}^{\infty} a_n\, (is)^n 
= 1 + a_1(x_2,x_1)\, (is) + a_2(x_2,x_1)\, (is)^2 + \ldots,
  \label{prd140}
 \end{equation}
 where $a_0=1$ and the next two coefficients are given by:
 \begin{equation}
a_1(x,x)  =  \frac{1}{6}\ R(x)
  \label{prd143}
 \end{equation}
 and
 \begin{equation}
 a_2(x,x) 
= \frac{1}{180}\left[
g^{\mu \nu} R_{;\mu;\nu}(x)  
+   R_{\mu\nu\lambda\rho} R^{\mu\nu\lambda\rho} 
-  R_{\mu\nu}R^{\mu\nu}\right].
  \label{prd144}
 \end{equation}
 Using these expressions in \eq{seven} we can determine how the background curvature affects the path measure in an arbitrary spacetime. We find, after some simple algebra, the result:
 \begin{equation}
 \cN = \frac{\Delta^{1/2}(x_1,x_2)}{(4\pi)}\frac{1}{\pi^k}\ \sum_{n=0}^\infty \frac{a_n}{4^n} \frac{\Gamma(k-n)}{(\sigma^2 - \rho^2 + i\epsilon)^{k-n}}
  \label{mon2}
 \end{equation}
 Obviously, the $n=0$ term gives back the flat spacetime limit (with $\Delta=1$) found earlier in \eq{apseven}, as it should. 
 
 The corresponding expression for the measure of closed loops of length $\sigma$  is obtained by setting $x_2=x_1=x$ and noting that, in this limit, we can take $\Delta=1$. We then get:
 \begin{equation}
 C(x,\sigma) = \frac{1}{(4\pi)}\frac{1}{\pi^k}\ \sum_{n=0}^\infty \frac{a_n}{4^n} \frac{\Gamma(k-n)}{(\sigma^2 + i\epsilon)^{k-n}}
  \label{tpmon2}
 \end{equation}
 This result allows us to obtain the gravitational action from the path measure, which we will explore in Sec. \ref{sec:three}.
 
 In some of the curved spacetimes, in which the heat kernel is known in explicit from, we can also obtain the expressions for $\cN$. I will mention one specific example, postponing its detailed discussion of a future publication. Consider the Einstein static universe with radius $H^{-1}$ in $D=4$. The zero-mass kernel for this geometry is well-known \cite{compores} and is given by:
 \begin{equation}
  \cK =\left(\frac{i}{16\pi^2 s^2}\right)\Delta^{1/2}\exp \left[-\frac{i\rho^2(x_1,x_2)}{4s} +isH^2\right]
 \end{equation} 
 Here $\rho^2=(t_2-t_1)^2-R^2(\mathbf{x}_1,\mathbf{x}_2)$ with $R(\mathbf{x}_1,\mathbf{x}_2)$ being the geodesic distance on $S^3$. The $\Delta$ is the Van-Vleck determinant given by: $\Delta=[HR/\sinh(HR)]$. Straightforward calculation using \eq{seven} 
 shows that the integral reduces to $K_{3/2}(z)$ which can be expressed in terms of elementary functions. This leads to the path density:
 \begin{eqnarray}
  \cN&=&\frac{1}{8\pi^2}\frac{1}{(\sigma^2-\rho^2)^{3/2}}
  \left[(1+H(\sigma^2-\rho^2)^{1/2})\exp(- H(\sigma^2-\rho^2)^{1/2})\right]\nonumber\\
  &=&\cN_{flat}\left[(1+H(\sigma^2-\rho^2)^{1/2})\exp(- H(\sigma^2-\rho^2)^{1/2})\right]
 \end{eqnarray} 
 where $\sigma^2$ is to be interpreted as the limit of $\sigma^2+i\epsilon$. We see that in this case the effect of curvature is to multiply the flat spacetime result in \eq{apseven} by the factor in the square brackets --- which reduces to unity when $H\to0$, as it should.

 \subsection{Effective Lagrangian and the measure for closed loops}
 
 It is straightforward to invert \eq{seven} and express $\cK$ in terms of $\cN$.  To do this, we first note that \eq{seven} implies that $\cN$ depends on $\sigma$ only through $\sigma^2$.  We set $\sigma^2 =q$ and treat  $\cN$  as a function of $q$ along the entire real line, analytically continuing to negative values of $q$. Then, it is easy to invert \eq{seven} and obtain: 
 \begin{equation}
  \cK = \left(\frac{i}{16\pi}\right)^{1/2} \frac{1}{s^{3/2}} \int_{-\infty}^\infty dq\ N(x_2,x_1; \sigma^2=q)\ e^{iq/4s}
  \label{eight}
 \end{equation}
 The pair of equations \eq{seven}, \eq{eight} tells us that $\cN$ and $\cK$ contain the same amount of geometrical information. (These relations are not exactly Fourier transforms but are closer to Laplace transforms; the  nature of this `Gaussian' transform is clarified in Appendix 2).
 
 The effective Lagrangian in the background spacetime, obtained by integrating out the scalar field is related to the kernel $\cK$ at the coincidence limit. 
 Using \eq{eight} to express $\cK$ in terms of $\cN$, we can relate the effective Lagrangian to the measure for the closed loops of length  $\sigma$ at $x$, given by \eq{newseven}. This is most easily done in the Euclidean sector as follows:
 In the Euclidean sector, we have
 \begin{equation}
  L^E_{\rm eff}(x;m) = - \frac{1}{2} \int_0^\infty \frac{ds}{s} \, e^{-m^2 s} \ K_0^E(x,x;s); \quad G^E(x,x;m) = \int_0^\infty K_0^E(x,x;s) ds
  \label{tp2}
 \end{equation} 
 These are formal expressions because, as they stand, they are divergent near $s=0$. (As is well-known, the effective Lagrangian has to be regularized to extract meaningful quantities.) We can think of these expressions as defined with a cut-off to the integral near $s=0$. With such a prescription, we have  $G^E = 2(\partial L^E_{\rm eff} / \partial m^2)$. Integrating, we have:
 \begin{equation}
  L^E_{\rm eff}(x;m)=\int_\infty^m G^E(x,x;\bar m)\bar m d\bar m
  \label{tp3}
 \end{equation} 
 Using \eq{six} and \eq{newseven}, we also have
 \begin{equation}
  G^E(x,x;m) = \int_{-\infty}^\infty d\sigma\ e^{-im\sigma}\ C(x;\sigma)
  \label{tp4}
 \end{equation} 
 Substituting \eq{tp4} into \eq{tp3} we easily get:
 \begin{equation}
  L^E_{\rm eff}(x;m) = - \int_{-\infty}^\infty \frac{d\sigma}{\sigma^2} \ C(x;\sigma)\, (1+im\sigma)\, e^{-im\sigma}
  \label{tp5}
 \end{equation} 
 with $\sigma$ interpreted as $\sigma - i\epsilon$. In the massless case, the effective Lagrangian simplifies to:
 \begin{equation}
   L^E_{\rm eff}(x;0) = - \int_{-\infty}^\infty \frac{d\sigma}{\sigma^2} \ C(x;\sigma) = -2 \int_0^\infty \frac{d\sigma}{\sigma^2} \ C(x;\sigma)
  \label{tp6}
 \end{equation} 
 Substituting \eq{tpmon2} into \eq{tp5} or \eq{tp6}, we can get the usual expansion of the effective Lagrangian in terms of the Schwinger-DeWitt coefficients. The first three terms, involving $a_0=1,a_1$ and $a_2$ will lead --- when evaluated with a cut-off $\Lambda$ --- to diverging contributions diverging as $\Lambda^{-4},\Lambda^{-2}$ and $\ln\Lambda$ respectively, just as in the usual analysis. This is one of the reasons why we cannot really obtain an induced gravity action (proportional to $R$) from the $a_1$ term because its coefficient diverges. However, as I will show in Sec. \ref{sec:three}, the situation is better when we attempt it with $C(x,\sigma)$; it is possible to obtain the Einstein-Hilbert action by using the distortion of $C(x,\sigma)$ by the background geometry directly.

\subsection{Aside: Effective Lagrangian in the electromagnetic case and the holonomy}
 
 So far we have been studying a scalar field in a given gravitational background such that the propagator in \eq{one} is related to the kernel in \eq{two} with $\Box $ being the Laplacian in the background spacetime with metric $g_{ab}$. All the above relations, especially the relation between the effective Lagrangian and the measure for closed loops, have a simple generalization even when an electromagnetic field is present in the background. When an electromagnetic field, described by a vector potential $A_j$ is present, \eq{one} and \eq{two} still hold with the zero mass kernel $K_0$ determined by \eq{two} with the replacement 
 \begin{equation}
 \Box \to -(i\partial_j - q A_j)^2
  \label{x1}
 \end{equation} 
 On the other hand, the path integral representation of the propagator in \eq{three} is now replaced by 
 \begin{equation}
 G(x_2,x_1; m) = \sum_{\rm paths} \exp \left[ -im\ell - i q \int_{x_1}^{x_2} dx^j\ A_j\right]
  \label{x2}
 \end{equation}
 Proceeding exactly as before, we can now define a quantity $\cN$ which will be given (instead of the expression in \eq{five}) by 
 \begin{equation}
 N(x_2,x_1;m) =  \sum_{\rm paths}^\sigma \exp \left[ - i q \int_{x_1}^{x_2} dx^j\ A_j\right]
  \label{x3}
 \end{equation} 
 where the superscript $\sigma$ on the summation symbol in the right hand side  reminds us that the sum is over all paths of a given length $\sigma$.  This expression, however, lacks the elegance of the one in \eq{five} because it is no longer a purely geometric construct. But when we study closed paths with the end points identified, $x_2=x_1 =x$, the situation gets a little better. In this case, the expression in \eq{newseven} gets replaced by 
 \begin{equation}
 C(x,m) =  \sum_{\rm paths}^\sigma \exp  - i q \oint_{\mathcal{C}} dx^j\ A_j
  \label{x4}
 \end{equation} 
 where the line integral is over a closed path $\mathcal{C}$ containing the point $x$ and having a length $\sigma$. This quantity
 \begin{equation}
 \oint_{\mathcal{C}} dx^j\ A_j = \int_{\mathcal{S}} d\Sigma^{ij}\, F_{ij} \equiv \Omega(\sigma; x)
  \label{x5}
 \end{equation} 
 is the flux of electromagnetic field over a surface $\mathcal{S}$ with $\mathcal{C}$ as a boundary and the $\Omega (\sigma; x)$ is closely related to the holonomy of the gauge field with the restriction that we are now considering closed paths of a given length $\sigma$. 
 
 The previous analysis of relating the effective Lagrangian to $C(x,\sigma)$ again goes through in a straightforward manner. As an example, consider the case of a slowly varying electromagnetic field and define $a,b$ by the relations 
 \begin{equation}
 E^2 - B^2 = a^2-b^2; \qquad \bm{E\cdot B} = ab
  \label{x6}
 \end{equation} 
 It is known that the Euclidean effective Lagrangian is then given by
 \begin{equation}
  L_{\rm eff}^E = \frac{1}{16\pi^2}\int_0^\infty \frac{ds}{s} \ e^{-m^2 s}\ P(s); \qquad P(s) \equiv\frac{qb}{\sinh qbs}\, \frac{qa}{\sin qas}
  \label{x7}
 \end{equation} 
 while the coincidence limit of the propagator is  given by the integral:
 \begin{equation}
 G^E(x,m) = 2 \frac{\partial L_{\rm eff}}{\partial m^2} = - \frac{1}{8\pi^2} \int_0^\infty ds\ P(s) \ e^{-m^2s}
  \label{x8}
 \end{equation} 
 From \eq{newseven} we find a simple expression for the measure for closed loops in the presence of an electromagnetic field:
 \begin{equation}
C(x,\sigma) = - \int_0^\infty \frac{ds}{16\pi^2} \frab{1}{\pi s}^{1/2} \ P(s) \, \exp\left( - \frac{\sigma^2}{4s}\right)
  \label{x9}
 \end{equation} 
 While the effective Lagrangian as well as the coincidence limit of the propagator are divergent (due to the behaviour of the integrand near $s=0$), the $C(x,\sigma)$ is finite, thanks to the strong convergence factor $\exp-(\sigma^2/4s)$.
 I hope to return to discuss this quantity and its relationship to pair production etc. in a later publication.

 \section{Einstein-Hilbert action from closed loops} \label{sec:three}
 
 We will now explore more closely the distortion of  $C(\sigma;x)$, from its flat spacetime value, by the curvature  of the spacetime, in $D=4$.
 We will work in the Euclidean sector and begin by noting that, \eq{tp1} allows us to express $\cN$  as an integral involving $\cK$.
 Further,  the Schwinger kernel in an arbitrary spacetime has an expansion in terms of Schwinger-DeWitt coefficients \cite{sch-dew} which allows us to write (in the Euclidean sector)
 \begin{equation}
K_0(s; x_2,x_1) = \frac{\Delta^{1/2}(x_2,x_1)}{(4\pi s)^{D/2}}\, e^{-\rho^2(x_2,x_1)/4s} F(x_2,x_1;s)
  \label{mon1new}
 \end{equation} 
 where $\rho(x_2,x_1)$ is the geodesic distance between the two events, $\Delta(x_2,x_1)$ is the Van-Vleck determinant and
 $F$ has the asymptotic series expansion given by \eq{prd140} in the Euclidean sector.
 The first two Schwinger-DeWitt coefficients, in the coincidence limit, are:
 $
a_0=1; 
a_1(x,x)  =  R(x)/6
 $.
  Using these expressions in \eq{seven} and taking the coincidence limit, we can determine how the background curvature affects $C(\sigma;x)$. This will lead to \eq{tpmon2} but now in the $D=4$ Euclidean sector; so we do not need the $i\epsilon$ factor and can set $D=4, k=(1/2)(D-1)=3/2$.
  In the resulting series, the $n=0$ term gives the flat spacetime contribution which is independent of $g_{ab}$ and can be dropped. Further, in the coincidence limit $x_2=x_1$, both $\rho$ and $\Delta$ can be set to unity, A straightforward calculation now gives: 
 \begin{equation}
 C(x,\sigma) = \frac{1}{(4\pi)}\frac{1}{\pi^{3/2}}\ \sum_{n=1}^\infty \frac{a_n}{4^n} \frac{\Gamma(\frac{3}{2}-n)}{\sigma^{3-2n}}
  = \frac{1}{(4\pi)}\frac{1}{\pi^{3/2}}\ \left[ \frac{a_1}{4} \frac{\Gamma(\frac{1}{2})}{\sigma}
 +\frac{a_2}{4^2} \Gamma(-\frac{1}{2})\sigma +\mathcal{O}(\sigma^3)\right]
 \label{tpmon2new}
 \end{equation}
 All the geometrical effects are contained in this series in which   \textit{only} the first term  diverges (as $1/\sigma$) in the $\sigma\to0$ limit. The corresponding measure on the length  contributed by these loops,
 $L(x;\sigma) \equiv \sigma C(x; \sigma)$, therefore, remains finite as $\sigma\to0$ limit. This leads to the remarkable result:
 \begin{equation}
  L(0; x) = \lim_{\sigma\to 0}\ L(\sigma,x) =  \frac{R(x)}{96\pi^2}
 \end{equation}  
 Since the limit  $\sigma\to0$ kills all higher order terms of Schwinger-DeWitt expansion, \textit{this expression is exact}. Integrating the result over a region of spacetime $\mathcal{V}$ we get:
 \begin{equation}
  A_{EH}\propto\lim_{\sigma\to 0}\ \int \sqrt{-g}d^4x [\sigma C(x,\sigma)]=\frac{1}{16\pi L_P^2}\int \sqrt{-g}d^4x R
  \label{agrav}
 \end{equation} 
 which is just the Einstein-Hilbert action. In the last expression we have restored the dimensions and set $\lambda=1/(6\pi)^{1/2}$ to match with standard definitions.
 Let me briefly sketch the implications of this result. 

At a technical level, the result is reminiscent of induced gravity models in which the Hilbert action is obtained from the effective Lagrangian for a scalar field  which, of course, can be related to the Schwinger Dewitt expansion etc \cite{adler}. However, even at the technical level the current approach is more elegant and precise in the sense that it picks out exactly what is needed through the limiting procedure in \eq{agrav}.
 \textit{I stress that this relation is exact} and is not just the first term in a series. (I have dropped the $n=0$ term which is divergent but independent of $g_{ab}$. Retaining it would have led, as usual, to a divergent cosmological constant in \eq{agrav} which anyway would have to renormalized to a finite/zero value.)
 
 What is more interesting is the deeper conceptual aspect of the result, viz., quantum paths of \textit{zero length} pick out the Ricci scalar. A classical, geometrical path is continuous and smooth with a well-defined  tangent vector to the curve at each point. Since this is forbidden in quantum theory, smooth trajectories do not contribute to a path integral; that is, they form a set of zero measure. Paths which do contribute to the path integral, which I have been calling the quantum paths, are very different in character. They can indeed be ``counted'', in terms of a finite measure $\cN$ for the relativistic path integral,  such that it correctly reproduces the propagator through \eq{four}.  The corresponding quantum measure for the closed loops, $C(\sigma;x)$, leads to a finite limit for 
 $\sigma C(\sigma;x)$  when the loops are shrunk to zero size in a curved geometry --- which, of course, cannot happen with smooth, classical paths. Roughly speaking, with our quantum measure the effective number of closed loops of lengths $\sigma$ scales as $R(x)/\sigma$ when $\sigma \to 0$. Therefore the total length contained in \textit{infinitesimal} closed paths remains finite and is proportional to $L_P(L_P^2R)$.  This is a purely  quantum effect as can be seen by the $\hbar$ dependence through the Planck length. Thus, the underlying quantum nature of the paths, which occur in relativistic path integral,  allows us to probe the geometrical features of the spacetime, in the limit of zero length. 
 
 It has been noticed earlier  that the Ricci \textit{tensor} occurs in the description of the density of states of quantum geometry  at mesoscopic scales \cite{tpdensity}. The current work shows that the Ricci \textit{scalar} can also play a similar role when we probe the spacetime using the quantum paths.   Neither of these results have a classical analogue: In classical differential geometry curvature components are usually associated with area or volume rather than with lengths. We see that quantum geometry has a far richer structure  than its classical counterpart. These aspects deserve deeper investigation.

 \section{Evaluation of a class of path integrals}
 
 As we mentioned earlier, the Feynman propagator can be defined through a path integral based on an action  $\mathcal{A}=-m\ell$ which, classically, describes a relativistic particle. It will be interesting to generalize this action and study relativistic path integrals in which the amplitude $(\exp-im\ell)$ is modified to the form
 \begin{equation}
 e^{-im\ell} \Rightarrow A(m,\ell) \equiv M(\ell)\exp-imS(\ell) 
 \label{mell}
 \end{equation}
 This is equivalent to studying an action, modified from $-m\ell$ to  $-mS(\ell)$,  with the path integral measure modified by a factor $M(\ell)$ relative to the standard measure.  Note that both the functions $M$ and $S$ are independent of $m$. One motivation for studying such a system arises from the fact that quantum gravitational effects could introduce a zero-point length into the spacetime \cite{zpl} changing $\ell$ to some function $S(\ell)$. Such an effect could also, generically, change the measure by a factor $M(\ell)$. (I will not discuss this motivation in detail here; see e.g., \cite{zpl}).
 
 Since we only know how to define (and calculate) path integrals for quadratic actions, the standard approach does not work for this case. (When $\mathcal{A}=-m\ell$   one can give meaning to path integral by lattice regularization but this procedure  does not generalize easily. There are also other issues in this approach; see \cite{tpnrqmpi}). It will be nice to have a general ansatz for evaluating such path integrals.  Needless to say, any technique to evaluate nontrivial path integrals is of interest on its own --- which is the second motivation for this study. I will show how this is indeed possible using $\cN$. 
 
 This is, in fact, fairly straightforward.
  We now want to evaluate the path integral:
 \begin{equation}
  G_{A(\ell)}= \sum_{\rm paths} A(m,\ell) =\sum_{\rm paths}M(\ell)\ \exp-imS(\ell)
 \end{equation} 
 Once again inserting the Dirac delta function we get:  
 \begin{equation}
  G_{A(\ell)} \equiv \sum_{\rm paths}  A(m,\ell) = \int_{-\infty}^\infty d\sigma \ A(m,\sigma)\, \sum_{\rm paths} \delta_D ( \sigma - \ell(x_2,x_1))
  \label{fourteen}
 \end{equation}
 which, on using our definition of $\cN$ reduces to  integral:
 \begin{equation}
 G_{A(\ell)}(x_2, x_1; m) = \int_{-\infty}^\infty d\sigma \ A(m,\sigma)\,\cN = \int_{-\infty}^\infty d\sigma \ e^{-iS(\sigma)} [M(\sigma) N(x_2,x_1;\sigma)]
  \label{fifteen}
 \end{equation}
\textit{Thus evaluating the path integral for an arbitrary functional of $\ell$ has been reduced to the task of evaluating an ordinary integral!} The last expression separates out the dependence on (i) the measure $M(\sigma)$ (which modifies $\cN$) and (ii) the new action $S(\sigma)$.
 
 One can gain some insight into this modification by re-expressing these results in terms of the Schwinger kernel. Let the kernel corresponding to the modified propagator be $K_{A(\ell)}(x_2,x_1,s)$ so that:
 \begin{equation}
 G_{A(\ell)}(x_2, x_1; m) = \int_0^\infty ds \ K_{A(\ell)} (x_2,x_1; s;m)
  \label{sixteen}
 \end{equation}
 Using \eq{seven} to express $\cN$ in terms of $\cK$ one can show that the new kernel, corresponding to $A(\ell)=M(\ell) \exp[-imS(\ell)]$ is related to the old zero-mass kernel for $[\exp -im\ell]$ by the relation:
 \begin{eqnarray}
 K_{A(\ell)} (x_2,x_1; s;m) &=&  K_0 (x_2,x_1; s) \frac{1}{(4\pi i s)^{1/2}} \int_{-\infty}^\infty d\sigma \ M(\sigma)e^{-im S(\sigma) +i\sigma^2/4s}\nonumber\\
 &\equiv& K_0 (x_2,x_1; s) \frac{1}{(4\pi i s)^{1/2}} I(s,m)
  \label{seventeen}
 \end{eqnarray}
 where we have defined
 \begin{equation}
 I(s,m) \equiv (4\pi i s)^{1/2} \frac{K_{A(\ell)}}{K_0} =\int_{-\infty}^\infty d\sigma M(\sigma) \, e^{-imS(\sigma)} \ e^{i\sigma^2/4s}\equiv \int_{-\infty}^\infty d\sigma A(m, \sigma) \ e^{i\sigma^2/4s}
\label{eighteen}
 \end{equation}
 We see that the two kernels are related by a factor which depends \textit{only} on $m$ and $s$ \textit{but not on the spatial coordinates}. In other words, the modification does not change the spatial dependence of the kernels which is a strong result and could not have been guessed without the use of $\cN$ to define the measure.

 One can ask whether we can accommodate any kind of modification --- that is any choice of $I(s,m)$ --- by choosing $M(\ell)$ and $S(\ell)$ appropriately. This is not possible in general. We can \textit{always} find a $A(\ell,m)$ which will lead to \textit{any} given $I(s,m)$. To determine  $A(\ell,m)$ corresponding to a 
 given $I(s,m)$ we only need to note that the function $I(s,m)$, defined through \eq{eighteen},
 is  the Gaussian transform of $A(m,\sigma)$. It can be inverted, in the general case, by using the results in Appendix 2. 
 But an such an $A(\ell,m)$, in general, cannot be expressed in the form $A(\ell,m) = M(\ell) \exp -imS(\ell)$ where the $m$ dependence is very specific. 
 
Let me illustrate this method, using an example, in which the desired modification of the kernel is of the form:
 \begin{equation}
 K_{A(\ell)}(x_2,x_1,s)= \cK e^{-im^2s}Q(s) 
 \label{modkernell}
 \end{equation}
 so that the mass dependence is of standard form and all the new physics is through the factor $Q(s)$. We will use the ansatz (which will be verified in the sequel) that  $A(m,\sigma)$ depends on $\sigma$ only through $\sigma^2$ so that $A(m,\sigma)=\bar A(m,\sigma^2)$. (The over-bar denotes the change in the functional dependence when we use $\sigma^2$ instead of $\sigma$.) From \eq{eighteen} we then have:
 \begin{equation}
  2\int_0^\infty d\sigma\ \bar A (m,\sigma^2) \ e^{i\sigma^2/4s} 
  = e^{-im^2 s } (4\pi i s)^{1/2} \ Q(s)
  \label{nineteen}
 \end{equation}
  This relation can be inverted  (see Appendix 2) to express 
 $A(m,\sigma)=\bar A(m,\sigma^2)$ in terms of $Q(s)$. We get
 \begin{equation}
  \frac{A(m,\sigma)}{|\sigma|} = \frac{\bar A(m,\sigma^2)}{\sqrt{\sigma^2}} = \frab{i}{\pi}^{1/2} \int_0^\infty dt\ e^{-i\sigma^2 t^2 - im^2/4t^2}\ Q \left(s= \frac{1}{4t^2}\right) + \text{c.c}
  \label{twenty}
 \end{equation}
 So, for any specific modification of the kernel given by \eq{modkernell}, determined by some function $Q(s)$, we can always find an action functional  $A(m,\sigma)$ such that the path integral leads to the correct modification.
 
 As a simple but important illustration, consider the case in which  $Q(s)=\exp \pm(i\lambda^2/4s)$ where $\lambda$ is a length scale. Such a modification will replace $(x-y)^2$ by $(x-y)^2\mp\lambda^2$ in the flat spacetime kernel, justifying the terminology `zero-point-length'\footnote{With our signature convention, $Q(s)=\exp +(i\lambda^2/4s)$ adds the zero-point-length to the spatial part while
 $Q(s)=\exp -(i\lambda^2/4s)$ adds a zero-point-time to the time interval. If we do everything in the Euclidean sector and analytically continue back, we end up adding it to the spatial length interval.} for $\lambda$. 
 Using the form
  \begin{equation}
 Q = e^{\pm i\lambda^2/4s}  = e^{\pm i\lambda^2 t^2}
  \label{twoone}
 \end{equation}
 in \eq{twenty} and performing the integral, we get
  \begin{equation}
  A(m, \sigma) = \frac{1}{2} \frab{\sigma^2}{\sigma^2\mp\lambda^2}^{1/2} \left[ e^{-im \sqrt{\sigma^2\mp\lambda^2}} \right] +\mathrm{c.c}
  \label{twotwo}
 \end{equation}
 We thus get the final result:\footnote{We have dropped the `c.c' term in \eq{twotwo} and multiplied the result by two, since both terms contribute the same amount, which is clear from the fact that the final result is invariant under $m\to -m$.}
 \begin{eqnarray}
 G(x_2,x_1)&=&\sum_{\rm paths}\frab{\ell^2}{\ell^2\mp \lambda^2}^{1/2}\ \exp -im \left( \ell^2 \mp \lambda^2\right)^{1/2} \nonumber\\
 &=&  i \int_0^\infty \frac{ds}{(4\pi^2 is)^{D/2}} \, e^{-im^2s} \exp -\frac{i}{4s}\left[(x_2 - x_1)^2 \mp \lambda^2\right]
  \label{fri1}
 \end{eqnarray} 
(The integral can be evaluated in terms of modified Bessel functions but its form is not very illuminating.) This is an interesting result  which tells us how to incorporate  the zero-point-length $\lambda$ into the path integral. In the left hand side, the propagator arises from a modified action; viz., the action has been modified from $-m\ell$ to $-m(\ell^2\mp\lambda^2)^{1/2}$ which takes into account the zero-point-length. We see that the net effect of this is just to change the interval $(x_2-x_1)^2$ to $(x_2-x_1)^2\mp\lambda^2$ in  the propagator. This seems quite reasonable because we have made the same change in both left and right hand sides. But note that this result is possible \textit{only when} we change the measure of the path integral by multiplying it by the factor $[\ell^2/(\ell^2+\lambda^2)]^{1/2}.$ Again one could not have guessed the result. 

The modified kernel and propagator have better UV behaviour and the propagator has a finite coincidence limit. (The properties of this propagator has been extensively explored in e.g \cite{tpduality} and I will not discuss them here.) The modification of the measure and the action in the path integral allows incorporating a fully Lorentz-invariant cut-off involving the zero-point length. We have shown that it can be obtained with well-defined measure for the path integral.\footnote{This fact can be confusing if you think of propagators as defined \textit{only} as two-point functions, arising in  unitary, local, Lorentz invariant \textit{field theories} rather than as defined by path integrals. You cannot get the propagator  in \eq{fri1} as a two-point function of a local, unitary, Lorentz invariant quantum field theory (see, e.g., page 127 of \cite{tpqft}). But it is rather unreasonable to expect a local, unitary,  quantum field theoretical description  to survive close to Planck length; so I am not bothered by this fact.}

The choice of measure used in \eq{fri1} is equivalent to assuming that the measure $\cN$ of paths of length $\sigma$ between two events is now modified from the result in \eq{apseven}  to the expression
\begin{eqnarray}
N_{\rm new} (x_2,x_1;\sigma) &=& \frac{1}{4\pi} \frac{\Gamma(k)}{\pi^k}\frab{\sigma^2}{\sigma^2\mp \lambda^2}^{1/2} \frac{1}{(\sigma^2 - x^2 +i\epsilon)^k}\nonumber\\
& \to  & \frac{1}{8\pi^2} \frab{\sigma^2}{\sigma^2\mp \lambda^2}^{1/2} \frac{1}{(\sigma^2 - x^2 +i\epsilon)^{3/2}}
 \label{fri2}
\end{eqnarray} 
where the last result is for $D=4, k=3/2$. The corresponding expression for  closed loops of length $\sigma$ is given by: 
 \begin{equation}
C_{\rm new} (\sigma) = \frac{1}{8\pi^2}\frab{\sigma^2}{\sigma^2\mp \lambda^2}^{1/2} \frac{1}{(\sigma^2  +i\epsilon)^{3/2}}
 \label{fri3}
\end{equation} 
If we ignore the $i\epsilon$ and the branch cuts, we get $C(\sigma)\propto (1/\lambda)\sigma^{-2}$ near $\sigma\to0$. So the number of loops of zero length diverges as $(1/\lambda) \sigma^{-2}$ compared to the divergence $\sigma^{-3}$ encountered in the absence of a zero-point-length.  The zero-point-length does not make everything finite, contrary to what one might have thought.

\section{Discussion}

The main purpose of this paper was to introduce and explore the concept of path density, $\cN$. Mathematically this can be thought of as a measure which allows \textit{functional} integration over paths to be replaced by \textit{ordinary} integration over path lengths. As I have shown, this feature by itself provides a powerful tool for computing a wide class of relativistic path integrals and exploring their properties. 

Conceptually one can think of $\cN$ as the (complex) amplitude for a quantum path of length $\sigma$ to exist between two events $x_1$ and $x_2$. This is a purely geometrical quantity dependent essentially on the background metric.  Though, algebraically, it contains exactly the same amount of information as the zero mass heat kernel, $\cK$, it is better suited to study microstructure of spacetime. One can translate  modifications of spacetime structure to the modification of $\cN$. In fact, the coincidence limit of the path density, $C(x,\sigma)$ has an intimate relation to Einstein-Hilbert action; this relation allows a new approach towards induced/emergent gravity models. 

Just like the heat kernel or the Feynman propagator, the path density also depends on the choice of vacuum state in the spacetime. One can again incorporate different choices of vacuum state by different choices for $\cN$ in a given spacetime. Again, this translates the idea of inequivalent quantization (like, for e.g., Rindler and inertial quantization) to the modification of path density. 

This work makes a case for using $\cN$ as an alternative descriptor of several aspects of quantum dynamics.  I hope to explore detailed applications of this approach in future publications.

\section*{Acknowledgement}

I thank Karthik Rajeev for several rounds of discussions. I thank Sumanta Chakraborty and Kinjalk Lochan for comments on the draft. My research  is partially supported by the J.C.Bose Fellowship of Department of Science and Technology, Government of India.

 \section*{Appendix 1: Mathematical details}\label{app:one}
 
 Let me outline how to obtain the relationships between the kernel and the path density and use it to compute the result in \eq{apseven}.
  We will do the Fourier transform with the factor $\exp(im\sigma)$ in both Lorentzian and Euclidean sectors. In the Lorentzian sector, the relation
  \begin{equation}
 \cG = \int_{-\infty}^\infty d\sigma\ \cN e^{-im\sigma} = \int_0^\infty ds\ e^{-im^2 s} \cK
   \label{26mar22}
  \end{equation}
  allows us to obtain
    \begin{eqnarray}
 N(\sigma)&=&\int_{-\infty}^\infty \frac{dm}{2\pi}\, e^{im\sigma} \int_0^\infty ds\ e^{-im^2 s} K_0(s) = \int_0^\infty  \frac{ds}{2\pi}\, K_0(s) \left( \frac{\pi}{is}\right)^{1/2} \, e^{i\sigma^2/4s}\nonumber\\
 &=& \frab{1}{4\pi i}^{1/2} \int_0^\infty \frac{ds}{s^{1/2}} K_0(s) e^{i\sigma^2/4s} 
   \label{26mar23}
  \end{eqnarray}
  We have not displayed the functional dependence on the coordinates to simplify the notation. 
  It is assumed that the integrand has a convergent factor $\exp(-\epsilon s)$ coming from the replacement $m^2\to m^2 -i\epsilon$.  One can also work out the same expression by relating the Euclidean quantities $N_E (\sigma) $ and $K^E_0(s)$, keeping the Fourier transform function as $\exp (im\sigma)$. We then get 
    \begin{eqnarray}
 N_E(\sigma) &=& \int_{-\infty}^\infty \frac{dm}{2\pi}\, e^{im\sigma}  \int_0^\infty ds\ e^{-m^2 s} K_0^E(s) = \int_0^\infty  \frac{ds}{2\pi}\, K_0^E(s) \left( \frac{\pi}{s}\right)^{1/2} \, e^{-\sigma^2/4s}\nonumber\\
 &=&  \frac{1}{(4\pi )^{1/2}} \int_0^\infty \frac{ds}{s^{1/2}} K_0^E(s) e^{-\sigma^2/4s} 
   \label{26mar24}
  \end{eqnarray}
  Let us consider next the inverse of the relation 
     \begin{equation}
  N(\sigma) = \frab{1}{4\pi i}^{1/2} \int_0^\infty \frac{ds}{s^{1/2}} \, e^{i\sigma^2/4s}\ K_0(s)
   \label{26mar25}
  \end{equation} 
  We note that the LHS depends only on $q\equiv \sigma^2$. We will treat the LHS as a function of $q$ with the same functional form holding for negative values of $q$ as well. That is, we take
  \begin{equation}
 N(q) = \frab{1}{4\pi i}^{1/2} \int_0^\infty \frac{ds}{s^{1/2}} \, K_0(s)\  e^{(i/4s)\, q}
   \label{26mar26}
  \end{equation}
  where $-\infty < q < +\infty$. Multiplying both sides by $e^{-ipq}$ and integrating over $dq/2\pi$ from $q=-\infty$ to $q=+\infty$, we obtain a Dirac delta function on the variable $p-(i/4s)$. This is same as a Dirac delta function on $s-(1/4p)$ multiplied by the Jacobian $4s^2$. This allows us to obtain the relation with, $s>0$   
  \begin{equation}
 \int_{-\infty}^\infty\frac{dq}{2\pi}\, N(q) e^{-iq/4s} = \frab{4}{i\pi}^{1/2} K_0(s) \, s^{3/2}
   \label{26mar27}
  \end{equation}
  which, in turn, leads to an expression for $K_0(s)$ in terms  of $N(q = \sigma^2)$:
  \begin{equation}
  K_0(s) = \frac{1}{s^{3/2}}\frab{i\pi}{16\pi^2}^{1/2}\int_{-\infty}^\infty dq\ N(q) e^{-iq/4s} 
   \label{26mar28}
  \end{equation}
  
  The flat spacetime results in \eq{apseven} can be easily obtained using \eq{26mar23} or \eq{26mar24}. Instead of doing that let me demonstrate the inverse relation, viz. that the measure in \eq{apseven} leads to the correct kernel. 
 To do this in the Lorentzian sector, let us start with a slightly more general   ansatz $N(q) = N_0 [-iq + iF(x_1,x_2)]^{-k}$ and see what is the form of the kernel which one obtains from this. Using the integral representation for the inverse power: 
  \begin{equation}
\frac{1}{t^k} = \frac{1}{\Gamma(k)} \, \int_0^\infty d\lambda \ \lambda^{k-1}\ e^{-\lambda t}
   \label{26mar36}
  \end{equation}
  we find that \eq{26mar28} leads to the result 
  \begin{equation}
   \cK = \frac{i}{(4\pi is)^{D/2}} \exp\left( - \frac{iF(x_1,x_2)}{4s}\right)
  \end{equation} 
  provided we choose the normalization constant  $N_0$ such that we get: 
  \begin{equation}
 N = \frac{1}{4\pi}\,  \frac{\Gamma(k)}{\pi^k}\, \frac{1}{(q-F(x_1,x_2) + i \epsilon)^k}
   \label{26mar37}
  \end{equation}
  In the final expression we have introduced the $i\epsilon$ factor in order to ensure convergence of the integral.

  Carrying out the corresponding calculation in the Euclidean sector is slightly more complicated because we need to deal with inverse \textit{Laplace} transforms, since
   \eq{26mar24} involves a Laplace transform. Taking $\sigma^2 =q$, $s=1/4\lambda$ and writing $\mathcal{F}(\lambda) \equiv K_0[s = 1/4\lambda]/\lambda^{3/2}$, \eq{26mar24} becomes
  \begin{equation}
 N^E (q) = \frab{1}{16\pi}^{1/2} \int_0^\infty d\lambda\ \mathcal{F} (\lambda)\, e^{-\lambda q}
   \label{26mar29}
  \end{equation}
  We thus have the Laplace transform and inverse transform relations, given by 
  \begin{eqnarray}
 \left[ \mathcal{L}^{-1} \, N^E(q)\right] \, (\lambda) &=& \frac{1}{\sqrt{16 \pi}}\ \mathcal{F}(\lambda)\nonumber\\
 K_0(\lambda = 1/4s) &=& \frac{\sqrt{16\pi}}{(4s)^{3/2}} \, \left[ \mathcal{L}^{-1} \, N^E(q)\right] \, (\lambda)
   \label{26mar30}
  \end{eqnarray}
  Let us again try this out with ansatz $N(q) = N_0 [q+F(x_2,x_1)]^{-k}$ written in the integral representation in \eq{26mar36}. We use the inverse transform relation given by
  \begin{equation}
  \mathcal{L}\left[ \lambda^n e^{-\lambda a}\right] (q) = \frac{n!}{(q+a)^{n+1}}
   \label{26mar31}
  \end{equation}
  and obtain 
  \begin{equation}
 \cK = \frac{(16\pi)^{1/2} N_0}{\Gamma(k)} \ \frab{1}{4s}^{k+1/2} \, e^{-F(x_1,x_2)/4s}
   \label{26mar32}
  \end{equation}
  It is easy to verify that for a choice 
  \begin{equation}
 N_0 = \frac{\Gamma(k)}{4\pi^{k+1}} = \frac{1}{4\pi} \, \frac{\Gamma(k)}{\pi^k}
   \label{26mar33}
  \end{equation}
  the resulting path measure 
  \begin{equation}
N_E(\sigma) = \frac{1}{4\pi} \, \frac{\Gamma(k)}{\pi^k}\, \frac{1}{[\sigma^2+F(x_1,x_2)]^k}
   \label{26mar34}
  \end{equation}
 leads to the Kernel
  \begin{equation}
 K_0^E(x_1,x_2;s)  = \frab{1}{4\pi s}^{D/2} \ \exp - \frac{F(x_1,x_2)}{4s}\ ; \qquad k= \frac{1}{2} (D-1)
   \label{26mar35}
  \end{equation}
  which is indeed the correct zero mass Schwinger kernel in $D$ dimensions if we take $F(x_1,x_2)=x^2$.

 \section*{Appendix 2: The Gaussian transform}
 
 Consider two functions $F(x)$ and $f(k)$ related by:
 \begin{equation}
  \int_0^\infty dx \ F(x) \ e^{ikx^2} \equiv f(k)
  \label{aa}
 \end{equation}
 We will assume that $F$ depends on $x$ only through $x^2$ and will write $F(x)=\bar F(x^2)$. The over-bar indicates that the functional form is different when expressed in terms of $x^2$ but the numerical values of the two sides are the same at any given $x$. Multiplying both sides of \eq{aa} by $\exp-iky^2$ and integrating over all $k$ we get:
 \begin{eqnarray}
 \int_{-\infty}^\infty \frac{dk}{2\pi}\ f(k)\ e^{-iky^2} &=& \int_0^\infty dx \ F(x) \, \delta_D[x^2-y^2]\nonumber\\
 &=& \int_0^\infty \frac{dx^2}{2 \sqrt{x^2}} \ \bar F(x^2) \, \delta_D[x^2-y^2] = \frac{\bar F(y^2)}{2\sqrt{y^2}} = \frac{F(y)}{2|y|}
  \label{bb}
 \end{eqnarray}
 which is the inverse transform to \eq{aa}. It will be interesting to examine how this $F(y)$, when inserted into \eq{aa}, does indeed lead to $f(k)$. The verification of this result proceeds as follows:
 \begin{eqnarray}
 \int_0^\infty dz \ F(z) \ e^{ikz^2} &=& \int_0^\infty \frac{dz^2}{2\sqrt{z^2}} \, \bar F(z^2) \ e^{ikz^2}
 = \int_0^\infty dz^2  \ e^{ikz^2}\int_{-\infty}^\infty \frac{dp}{2\pi}\ f(p)\ e^{-ipz^2}\nonumber\\
 &=& \int_{-\infty}^\infty \frac{dp}{2\pi}\ f(p)\int_0^\infty dz^2  \ e^{iz^2(k-p+i\epsilon)}\nonumber\\
 &=& \int_{-\infty}^\infty \frac{dp}{2\pi}\ \frac{i f(p)}{(k-p+i\epsilon)} = \int_{-\infty}^\infty \frac{dp}{2\pi i}\ \frac{f(p)}{(p-k-i\epsilon)} = f(k)
  \label{cc}
 \end{eqnarray}
 In the penultimate step we have introduced the appropriate $i\epsilon$ to ensure convergence and the last result follows from Cauchy's theorem assuming $f(p)$ has the required boundary behaviour. It is in this sense that the Gaussian transform in \eq{aa} and its inverse transform in \eq{bb} are to be interpreted.


 \end{document}